# REISSNER EXTERIOR AND INTERIOR

Rainer Burghardt[*]

Contents

The Reissner-Nordström metric is re-examined and supplemented with an interior solution. Both metrics are embedded in a 5-dimensional flat space.

---

[*] e-mail: arg@aon.at, home page: http://arg.or.at/

# 1. INTRODUCTION

Schwarzschild [1] has found the first solution of the Einstein field equations, which was published in 1916. Reissner and Nordström [2] have extended this solution to the field of electrically charged matter. It can be easily shown that the spacelike part of the exterior metric can be explained by a surface similar to Flamm's paraboloid. Covering the 'hole' of this surface with the cap of a sphere and treating the timelike part of the metric in the same way as we have done for the Schwarzschild interior [3], we obtain an interior for the Reissner-Nordström metric. This solution reduces to the Schwarzschild interior solution by putting the charge zero.

In Sec. 2, we re-examine the Reissner-Nordström solution. We perform a [3+1]-decomposition of the field equations and describe the forces acting on the observers in the exterior field. In Sec. 3, we construct an interior for the Reissner-Nordström metric, based on the geometrical methods used for the Schwarzschild interior solution. In Sec. 4, we examine the field equations of the interior solution, in Sec. 5 we discuss the embedding in a higher dimensional flat space, and we apply the theory of surfaces to this solution.

# 2. THE EXTERIOR SOLUTION

It is widely accepted that the electromagnetic field has the same influence on gravitation as the matter has. Reissner and Nordström have found a solution of the Einstein field equations where the stress-energy tensor is made up with the electric field strengths, and the electric field is the consequence of a charge term in the metric. In quasi-polar co-ordinates with e as the electric charge in geometrical units this metric reads as

$$ds^2 = \alpha^2 dr^2 + r^2 d\vartheta^2 + r^2 \sin^2\vartheta\, d\varphi^2 + a^2 dit^2$$
$$a^2 = 1 - \frac{2M}{r} + \frac{e^2}{r^2}, \quad \alpha = a^{-1} \qquad (2.1)$$

With the unit vectors

$$b_m = \{0,1,0,0\}, \quad c_m = \{0,0,1,0\}, \quad u_m = \{0,0,0,1\} \qquad (2.2)$$

we are able to decompose the Ricci rotation coefficients into

$$A_{mn}{}^s = \left[b_m B_n b^s - b_m b_n B^s\right] + \left[c_m C_n c^s - c_m c_n C^s\right] - \left[u_m G_n u^s - u_m u_n G^s\right]. \qquad (2.3)$$

B, C, and G are the field strengths. G consists of two parts, a gravitational and an electric one:

$$G_m = E_m + \mathcal{E}_m. \qquad (2.4)$$

The field strengths have the components

$$B_m = \left\{\frac{a}{r},0,0,0\right\}, \quad C_m = \left\{\frac{a}{r},\frac{1}{r}\cot\vartheta,0,0\right\}, \quad E_m = \left\{-\alpha\frac{M}{r^2},0,0,0\right\}, \quad \mathcal{E}_m = \left\{\alpha\frac{e^2}{r^3},0,0,0\right\}. \quad (2.5)$$

Particularly, $E_m$ is the force of gravity, formally identical with the force of gravity in the Schwarzschild field, and $\mathcal{E}_m$ is a new contribution to the field strengths. With these quantities the Ricci reads as

$$R_{mn} = -\left[\underset{2}{B_{n\|m}} + B_n B_m\right] - b_n b_m \left[\underset{2}{B^s_{\|s}} + B^s B_s\right]$$

$$-\left[\underset{3}{C_{n\|m}} + C_n C_m\right] - c_n c_m \left[\underset{3}{C^s_{\|s}} + C^s C_s\right] \quad (2.6)$$

$$+\left[\underset{4}{G_{n\|m}} - G_n G_m\right] + u_n u_m \left[\underset{4}{G^s_{\|s}} - G^s G_s\right]$$

wherein the graded derivatives [4] are used. As $R = 0$ we obtain by evaluating (2.6) and rescaling $e^2 \to \dfrac{\kappa}{2} e^2$ the field equations

$$R_{mn} - \frac{1}{2} g_{mn} R = -\kappa T_{mn}$$
$$T_{mn} = \mathcal{F}_m{}^s \mathcal{F}_{ns} - \frac{1}{4} g_{mn} \mathcal{F}^{rs} \mathcal{F}_{rs} \quad . \quad (2.7)$$

The electric field has only one component pointing into the radial direction of the model

$$\mathcal{F}_{14} = \mathcal{F}_1 = \frac{e}{r^2} \quad (2.8)$$

and satisfies the Maxwell equations

$$\mathcal{F}^{mn}{}_{\|n} = 0, \quad \mathcal{F}_{\langle mn\|s\rangle} = 0. \quad (2.9)$$

The electric stresses and the energy are conserved

$$T^{mn}{}_{\|n} = 0 \quad . \quad (2.10)$$

Evaluation of the last bracket of (2.6) shows that

$$\mathcal{E}^s{}_{\|s} = -\frac{\kappa}{2} \mathcal{F}^s \mathcal{F}_s \quad . \quad (2.11)$$

If this model is not a mere mathematical exercise one has to accept a repulsive electric short-range force

$$\mathcal{E}_s = \left\{\alpha \frac{\kappa}{2} \frac{e^2}{r^3}, 0, 0, 0\right\}, \quad (2.12)$$

quadratic in the charge and having the field energy as its source.

In order to prepare ourselves for matching an interior to this solution, we firstly have to search for a surface embedded in a higher dimensional flat space illustrating the geometry. Then we will proceed in the same way as we did for the Schwarzschild geometry [4]. As

$$v = -\sqrt{\frac{2M}{r} - \frac{e^2}{r^2}} \qquad (2.13)$$

is the velocity of a freely falling observer, we define the angle of ascent of a curve as

$$\sin\varepsilon = v \;, \qquad (2.14)$$

where the orientation of ε is cw. Then we find the quantity a introduced in (2.1) to be cos ε. Rotating this curve through the angle $\vartheta$ about the co-ordinate axis $\mathcal{R}^{0'}$ of the extra dimension we get a surface similar to Flamm's paraboloid of the Schwarzschild geometry. The same holds for the rotation through the angle φ. The timelike part of the surface can be understood by following the arguments we have published for the Schwarzschild geometry [4].

Since the radial force can be written as

$$G_1 = \frac{1}{\rho}\tan\varepsilon$$

we are able to calculate the curvature radius of the radial lines on the surface as

$$\rho = r^2 \frac{\sqrt{2Mr - e^2}}{Mr - e^2} . \qquad (2.15)$$

Now we have provided all the quantities we need for matching an interior to the exterior metric.

# 3. THE INTERIOR SOLUTION

Through the last decades, several attempts have been made to round out the Reissner metric with an interior. Tiwani, Rao, and Kanakamedala [5] found an interior with the condition $g_{11}g_{44}= 1$, which obviously holds for the exterior Reissner, but does not match the interior Schwarzschild solution for e = 0. Kyle and Martin [6] found a rather complicated solution. They discussed the self-energy of the fields of charged matter. Wilson [7] modified this solution assuming a different value for the total charge. Cohen and Cohen [8] applied this solution to the special case of a charged thin shell and showed that the energy density of the electromagnetic field contributes to the mass. Boulware [9] studied the time development of thin shells. Graves and Brill [10] considered a possible oscillatory character of the Reissner-Nordström metric by examining the metric by Kruskal-like co-ordinates. Bekenstein [11] studied an ansatz for the stress-energy-tensor of charged matter

$$T_{mn} = -p'g_{mn} + \mu_0 u_m u_n + F_m{}^s F_{ns} - \frac{1}{4} g_{mn} F^{rs} F_{rs} \tag{3.1}$$

and tried to adjust p and $\mu_0$ so that the field equations are satisfied. Krori [12] made use of the same ansatz, and he found a solution free of singularities. Gautreau and Hoffman [13] studied the sources of Weyl-type electrovac fields. They obtained the parameters for the source with the junction condition for the exterior solution. Efinger [14] deduced the stability of a charged particle from the self-energy of the gravitation field.

We construct another interior solution with the help of embedding the geometry in a 5-dimensional flat space. This method has the advantage that the solution reduces to the Schwarzschild interior solution by putting the charge e = 0. However, the decomposition (3.1) of the matter part and the electric part of the stress-energy tensor cannot be performed in a satisfactory way. Gravitational and electric forces are geometrically soldered in the interior.

For the spacelike part of the interior, we choose the cap of a sphere with the metric on this cap

$$ds^2 = \mathcal{R}^2 d\eta^2 + \mathcal{R}^2 \sin^2\eta \, d\vartheta^2 + \mathcal{R}^2 \sin^2\eta \sin^2\vartheta \, d\varphi^2. \tag{3.2}$$

$\mathcal{R}$ is the radius and $\eta$ the polar angle of the sphere. $\eta_g$ will be the aperture angle of the cap. The spherical co-ordinates of the sphere are connected with the Cartesian coordinates of the flat embedding space by

$$\begin{aligned} \mathcal{R}^{3'} &= \mathcal{R} \sin\eta \sin\vartheta \sin\varphi \\ \mathcal{R}^{2'} &= \mathcal{R} \sin\eta \sin\vartheta \cos\varphi \\ \mathcal{R}^{1'} &= \mathcal{R} \sin\eta \cos\vartheta \\ \mathcal{R}^{0'} &= \mathcal{R} \cos\eta \end{aligned} \tag{3.3}$$

$\mathcal{R}^{0'}$ is the co-ordinate line of the extra dimension. Introducing the standard Schwarzschild co-ordinate

$$r = \mathcal{R} \sin\eta \tag{3.4}$$

we obtain instead of (3.2)

$$ds^2 = \frac{1}{1 - \frac{r^2}{\mathcal{R}^2}} dr^2 + r^2 d\vartheta^2 + r^2 \sin^2\vartheta \, d\varphi^2 . \tag{3.5}$$

The *embedding condition* is

$$\mathcal{R} = \mathcal{R}_g = \text{const.} . \tag{3.6}$$

At the boundary surface of the interior and the exterior solution the angle of ascent[1] of the radial curves on the surfaces have to be equal $\eta_g = -\varepsilon_g$. From (2.13), (2.14), and (3.4) we obtain the *junction condition*

$$\mathcal{R}_g = \frac{r_g^2}{\sqrt{2Mr_g - e^2}} . \tag{3.7}$$

To get the timelike part of the interior line element we proceed in the same manner as we have done to explain the Schwarzschild interior [3]. We supplement the metric (3.2) with a timelike element

$$ds^2 = \mathcal{R}^2 d\eta^2 + \mathcal{R}^2 \sin^2\eta \, d\vartheta^2 + \mathcal{R}^2 \sin^2\eta \sin^2\vartheta \, d\varphi^2 + a_T^2 dit^2$$

$$a_T = \left[ (\mathcal{R}_g + \rho_g)\cos\eta_g - \mathcal{R}\cos\eta \right] \frac{1}{\rho_g} = \frac{1}{2} \left[ (1 + 2\Phi_g^2)\cos\eta_g - \cos\eta \right] \Phi_g^{-2} . \tag{3.8}$$

$$\rho_g = 2\mathcal{R}_g \Phi_g^2 = r_g^2 \frac{\sqrt{2Mr_g - e^2}}{Mr_g - e^2} = \mathcal{R}_g \frac{2Mr_g - e^2}{Mr_g - e^2}, \quad 2\Phi_g^2 = \frac{2Mr_g - e^2}{Mr_g - e^2}$$

$\rho_g$ we obtain from (2.15). It is the radius of curvature of the exterior radial curves at the boundary surface. $\mathcal{R}_g + \rho_g$ is the line segment starting at the tail of the curvature vector $\vec{\rho}_g$ and ending at the symmetry axis of the surface, labeled by $\mathcal{R}^{0'}$. $(\mathcal{R}_g + \rho_g)\cos\eta_g$ is the projection of this line segment onto the co-ordinate line of the extra dimension $\mathcal{R}^{0'}$ and $\mathcal{R}\cos\eta$ the projection of the radius vector to an arbitrary point of the cap of the sphere onto the co-ordinate line of the extra dimension. Rotating these projections through an imaginary angle, we get two pseudo-circles. If we apply the embedding condition, (3.6) the enclosed ring sector is proportional to the flow of time

$$dt = \rho_g d\psi . \tag{3.9}$$

For e = 0 the metric reduces to the metric of the Schwarzschild interior solution.

## 4. THE FIELD EQUATIONS

If we evaluate the components of the 4-bein from the metric (3.8) and if we also evaluate the field strengths for (2.3) we get

$$B_m = \left\{ \frac{1}{\mathcal{R}}\cot\eta, 0, 0, 0 \right\}, \quad C_m = \left\{ \frac{1}{\mathcal{R}}\cot\eta, \frac{1}{\mathcal{R}\sin\eta}\cot\vartheta, 0, 0 \right\}, \quad G_m = \left\{ -\frac{1}{\rho_g a_T}\sin\eta, 0, 0, 0 \right\} . \tag{4.1}$$

In addition, we define extra components

---

[1] To simplify the calculations, the orientation of the angles of ascent η and ε are chosen to be ccw and cw.

$$M_0 = \frac{1}{\mathcal{R}}, \quad B_0 = \frac{1}{\mathcal{R}}, \quad C_0 = \frac{1}{\mathcal{R}}, \quad G_0 = \frac{1}{\rho_g a_T}\cos\eta, \tag{4.2}$$

which we will explain in the next Section. With these expressions the Ricci tensor reads as

$$\begin{aligned}R_{mn} =\ & m_m m_n (M_0 B_0 + M_0 C_0 - M_0 G_0) \\ & + b_m b_n (B_0 M_0 + B_0 C_0 - B_0 G_0) \\ & + c_m c_n (C_0 M_0 + C_0 B_0 - C_0 G_0) \\ & - u_m u_n (G_0 M_0 + G_0 B_0 + G_0 C_0)\end{aligned}. \tag{4.3}$$

Defining also the quantity

$$\mathcal{P} = -\frac{\mathcal{R}\cos\eta}{\rho_g a_T} = -\frac{1}{\left(\frac{\mathcal{R}_g}{\mathcal{R}} + \frac{\rho_g}{\mathcal{R}}\right)\frac{\cos\eta_g}{\cos\eta} - 1} \tag{4.4}$$

we get from (4.3) with

$$R_{mn} - \frac{1}{2}g_{mn}R = -\kappa T_{mn}$$

the stress-energy tensor

$$T_{mn} = \begin{pmatrix} -p & & & \\ & -p & & \\ & & -p & \\ & & & \mu_0 \end{pmatrix}. \tag{4.5}$$

The hydrostatic pressure and the energy density of the source are

$$\kappa p = -(1+2\mathcal{P})\frac{1}{\mathcal{R}^2}, \quad \mu_0 = \frac{3}{\mathcal{R}^2}. \tag{4.6}$$

The electric parts of the stress-energy tensor cannot be separated from the matter contribution. On the boundary surface one has

$$\kappa p_g = -F_1^{\ g} F_1^{\ g}. \tag{4.7}$$

For e = 0 one obtains the stress-energy tensor of the Schwarzschild interior. The conservation law reads as

$$p_{\|m} = (\mu_0 + p)G_m, \quad \dot{p} = 0, \quad \dot{\mu}_0 = 0. \tag{4.8}$$

In the subsequent Section, we will show that the timelike part of the metric (3.8) can also be explained by embedding techniques and we will deepen the geometrical meaning of the stress-energy tensor and the conservation law.

# 5. EMBEDDING THE INTERIOR SOLUTION

In the last Section we have introduced new quantities (4.2). We interpret them as the components of the geometrical quantities in an extra dimension and we will show that the interior Reissner-Nordström metric is the metric of a geometrical object embedded in a 5-dimensional flat space. We extend the ansatz (3.3) to

$$X^{3'} = (\mathcal{R}_g + \rho_g)\sin\eta_g \sin\vartheta \sin\varphi - \mathcal{R}\sin\eta \sin\vartheta \sin\varphi$$
$$X^{2'} = (\mathcal{R}_g + \rho_g)\sin\eta_g \sin\vartheta \cos\varphi - \mathcal{R}\sin\eta \sin\vartheta \cos\varphi$$
$$X^{1'} = (\mathcal{R}_g + \rho_g)\sin\eta_g \cos\vartheta \quad - \mathcal{R}\sin\eta \cos\vartheta \quad . \quad (5.1)$$
$$X^{0'} = (\mathcal{R}_g + \rho_g)\cos\eta_g \cos\psi \quad - \mathcal{R}\cos\eta \cos\psi$$
$$X^{4'} = (\mathcal{R}_g + \rho_g)\cos\eta_g \sin\psi \quad - \mathcal{R}\cos\eta \sin\psi$$

Evidently both columns of (5.1) are related to a pseudo-hypersphere embedded in a 5-dimensional flat space. To perform the dimensional reduction we cut off the first three lines of the first column on the right side of (5.1) and we use the embedding condition (3.6). By differentiating the remaining expressions we can deduce the line element (3.8). We emphasize that the line element can only be understood by embedding two surfaces in a 5-dimmensinal flat space. However, it turns out that five dimensions are sufficient. The curvature quantities of the geometry we present by using a spherical co-ordinate system $\{\mathcal{R}, \eta, \vartheta, \varphi, \psi\}$:

$$A_{ab}{}^c = M_{ab}{}^c + B_{ab}{}^c + C_{ab}{}^c + G_{ab}{}^c$$
$$M_{ab}{}^c = m_a M_b m^c - m_a m_b M^c, \quad B_{ab}{}^c = b_a B_b b^c - b_a b_b B^c, \quad (5.2)$$
$$C_{ab}{}^c = c_a C_b c^c - c_a c_b C^c, \quad G_{ab}{}^c = -\left[u_a G_b b^c - u_a u_b G^c\right]$$

with a = 0, 1, …, 4, $m_a = \{0,1,0,0,0\}$. The components of the quantities M, B, C, G we read from (4.1) and (4.2). The flat 5-dimensional field equations

$$R_{ab} \equiv 0$$

decouple into the curvature equations

$$M_{a\underset{1}{|||}b} + M_a M_b = 0, \quad B_{a\underset{2}{|||}b} + B_a B_b = 0, \quad C_{a\underset{3}{|||}b} + C_a C_b = 0, \quad G_{a\underset{4}{|||}b} - G_a G_b = 0$$
$$M^c{}_{\underset{1}{|||}c} + M^c M_c = 0, \quad B^c{}_{\underset{2}{|||}c} + B^c B_c = 0, \quad C^c{}_{\underset{3}{|||}c} + C^c C_c = 0, \quad G^c{}_{\underset{4}{|||}c} - G^c G_c = 0 \quad . \quad (5.3)$$

The graded derivatives are used again

$$M_{a\underset{1}{|||}b} = M_{a|b}, \quad B_{a\underset{2}{|||}b} = B_{a|b} - M_{ba}{}^c B_c, \quad C_{a\underset{3}{|||}b} = C_{a|b} - M_{ba}{}^c C_c - B_{ba}{}^c C_c$$
$$G_{a\underset{4}{|||}b} = G_{a|b} - M_{ba}{}^c G_c - B_{ba}{}^c G_c - C_{ba}{}^c G_c \quad . \quad (5.4)$$

We rename the quantities (4.2) as

$$A_{11} = M_0, \quad A_{22} = B_0, \quad A_{33} = C_0, \quad A_{44} = -G_0$$
$$A_{11} = A_{22} = A_{33} = \frac{1}{\mathcal{R}}, \quad A_{44} = \frac{p}{\mathcal{R}} \qquad (5.5)$$

and we interpret them as the generalized second fundamental forms $A_{mn}$ of the geometry. They are components of the 5-dimensional connexion coefficients. Isolating them on the right side of the field equations, we obtain the stress-energy tensor

$$\kappa T_{mn} = A_m{}^s A_{ns} - A_{mn} A_s{}^s - \frac{1}{2} g_{mn} \left[ A^{rs} A_{rs} - A_r{}^r A_s{}^s \right] . \qquad (5.6)$$

The conservation law can be written as

$$\kappa T_m{}^n{}_{||n} = 2 A_{<m}{}^s A_{[s}{}^n{}_{||n]>} . \qquad (5.7)$$

It is satisfied by the Codazzi equation

$$A_{[m}{}^n{}_{||n]} = 0 \qquad (5.8)$$

Applying methods of the surface theory shows that it is not necessary to require the divergence of the right side of the Einstein field equations to vanish. It turns out to be a fundamental property of the geometrical structure. Matter and electric charge are described by the fields $A_{mn}$, and the field equations for these matter fields are the Codazzi equations.

## 6. SUMMARY

We have supplemented the Reissner-Nordström exterior metric with a metric for an interior solution. The metric has the same structure as the Schwarzschild interior metric and reduces to it by putting the charge zero. The Reissner-Nordström ansatz is a semi-unified theory of gravitation and electromagnetism. The electric stress-energy tensor has its origin in the curvature of space. However, the electric field strengths have to satisfy the Maxwell equations which are independent of the geometry.